\documentclass[a4paper,11pt]{article}

\usepackage{graphicx}
\usepackage{fullpage}
\usepackage{amsfonts}
\usepackage{amsmath}
\usepackage{subfigure}
\usepackage{setspace}
\usepackage{epigraph}
\usepackage{xcolor}
\usepackage{array,booktabs}
\usepackage{cprotect}
\usepackage{colortbl} 
\usepackage{hyperref}
\usepackage{upref}

\numberwithin{equation}{section}
\begin{document}

\title{To mask or not to mask: Modeling the potential for face mask use by the general public to curtail the COVID-19 pandemic}
\author{Steffen E. Eikenberry$^*$\footnote{email {seikenbe@asu.edu}}, Marina Mancuso$^*$, Enahoro Iboi$^*$, Tin Phan$^*$, Keenan Eikenberry$^*$, \\ Yang Kuang$^*$, Eric Kostelich$^*$, and Abba B. Gumel$^*$}
\date{$^*$Arizona State University, School of Mathematical and Statistical Sciences, Tempe, AZ, USA \\ \today}
\maketitle

\begin{abstract}
Face mask use by the general public for limiting the spread of the COVID-19 pandemic is controversial, though increasingly recommended, and the potential of this intervention is not well understood.  We develop a compartmental model for assessing the community-wide impact of mask use by the general, asymptomatic public, a portion of which may be asymptomatically infectious.  Model simulations, using data relevant to COVID-19 dynamics in the US states of New York and Washington, suggest that broad adoption of even relatively ineffective face masks may meaningfully reduce community transmission of COVID-19 and decrease peak hospitalizations and deaths.  Moreover, mask use decreases the effective transmission rate in nearly linear proportion to the product of mask effectiveness (as a fraction of potentially infectious contacts blocked) and coverage rate (as a fraction of the general population), while the impact on epidemiologic outcomes (death, hospitalizations) is highly nonlinear, indicating masks could synergize with other non-pharmaceutical measures.  Notably, masks are found to be useful with respect to both preventing illness in healthy persons \textit{and} preventing asymptomatic transmission.  Hypothetical mask adoption scenarios, for Washington and New York state, suggest that immediate near universal (80\%) adoption of moderately (50\%) effective masks could prevent on the order of 17--45\% of projected deaths over two months in New York, while decreasing the peak daily death rate by 34--58\%, absent other changes in epidemic dynamics.  Even very weak masks (20\% effective) can still be useful if the underlying transmission rate is relatively low or decreasing: In Washington, where baseline transmission is much less intense, 80\% adoption of such masks could reduce mortality by 24--65\% (and peak deaths 15--69\%), compared to 2--9\% mortality reduction in New York (peak death reduction 9--18\%). Our results suggest use of face masks by the general public is potentially of high value in curtailing community transmission and the burden of the pandemic.   The community-wide benefits are likely to be greatest when face masks are used in conjunction with other non-pharmaceutical practices (such as social-distancing), and when adoption is nearly universal (nation-wide) and compliance is high.
\end{abstract}

\section{Introduction}

Under the ongoing COVID-19 pandemic (caused by the SARS-CoV-2 coronavirus), recommendations and common practices regarding face mask use by the general public have varied greatly and are in rapid flux: Mask use by the public in public spaces has been controversial in the US, although as of April 3, 2020, the US Centers for Disease Control and Prevention (CDC) is recommending the public wear cloth masks. Public mask use is far more prevalent in many Asian countries, which have longer experience with novel coronavirus epidemics; public mask use may have been effective at limiting community spread during the 2003 SARS epidemic \cite{Wu2004,Lau2004}, and widespread mask use is a prominent feature of the relatively successful COVID-19 response in Taiwan \cite{Wang2020}, for example.  Masks have also been suggested as method for limiting community transmission by asymptomatic or at least clinically undetected carriers \cite{Chan2020}, who may be a major driver of transmissions of COVID-19 \cite{Li2020}. Various experimental studies suggest that masks may both protect the wearer from acquiring various infections \cite{Lai2012,Davies2013} or transmitting infection \cite{Dharmadhikari2012}.  Medical masks (i.e., surgical masks and N95 respirators) in healthcare workers appear to consistently protect against respiratory infection under metanalysis \cite{MacIntyre2017,Offeddu2017}, although clinical trials in the community have yielded more mixed results \cite{MacIntyre2009,Cowling2009,Canini2010}.  While medical-grade masks should be prioritized for healthcare providers, homemade cloth masks may still afford significant, although variable and generally lesser, protection \cite{Sande2008,Davies2013}, but clinical trials in the community remain lacking.

Given the flux in recommendations, and uncertainty surrounding the possible community-wide impact of mass face masks (especially homemade cloth masks) on COVID-19 transmission, we have developed a multi-group Kermack-McKendrick-type compartmental mathematical model, extending prior work geared towards modeling the COVID-19 pandemic (e.g. \cite{Li2020,Ferguson2020,Tang2020}), as well as models previously used to examine masks in a potential influenza pandemic \cite{Brienen2010,Tracht2010}.  This initial framework suggests that masks could be effective even if implemented as a singular intervention/mitigation strategy, but \textit{especially} in combination with other non-pharmaceutical interventions that decrease community transmission rates.

Whether masks can be useful, even in principle, depends on the mechanisms for transmission for SARS-CoV-2, which are likely a combination of droplet, contact, and possible airborne (aerosol) modes.  The traditional model for respiratory disease transmission posits infection via infectious droplets (generally 5--10 $\mu$m) that have a short lifetime in the air and infect the upper respiratory tract, or finer aerosols, which may remain in the air for many hours \cite{Leung2020}, with ongoing uncertainties in the relative importance of these modes (and in the conceptual model itself \cite{Bourouiba2020}) for SARS-CoV-2 transmission \cite{Han2020,Bourouiba2020}.  The WHO \cite{WHO2020} has stated that SARS-CoV-2 transmission is primarily via coarse respiratory droplets and contact routes.  An experimental study \cite{Doremalen2020} using a nebulizer found SARS-CoV-2 to remain viable in aerosols ($<$5 $\mu$m) for three hours (the study duration), but the clinical relevance of this setup is debatable \cite{WHO2020}.  One out of three symptomatic COVID-19 patients caused extensive environmental contamination in \cite{Ong2020}, including of air exhaust outlets, though the air itself tested negative.

Face masks can protect against both coarser droplet and finer aerosol transmission, though N95 respirators are more effective against finer aerosols, and may be superior in preventing droplet transmission as well \cite{MacIntyre2017}.  Metanalysis of studies in healthy healthcare providers (in whom most studies have been performed) indicated a strong protective value against clinical and respiratory virus infection for both surgical masks and N95 respirators \cite{Offeddu2017}.  Case control data from the 2003 SARS epidemic suggests a strong protective value to mask use by community members in public spaces, on the order of 70\% \cite{Wu2004,Lau2004}.

Experimental studies in both humans and manikins indicate that a range of mask provide at least some protective value against various infectious agents \cite{Sande2008,Davies2013,Driessche2015,Stockwell2018,Leung2020}.  Medical masks were potentially highly effective as both source control and primary prevention under tidally breathing and coughing conditions in manikin studies \cite{Lai2012,Patel2016}, with higher quality masks (e.g. N95 respirator vs. surgical mask) offering greater protection \cite{Patel2016}.  It is largely unknown to what degree homemade masks (typically made from cotton, teacloth, or other polyesther fibers) may protect against droplets/aerosols and viral transmission, but experimental results by Davies et al. \cite{Davies2013} suggest that while the homemade masks were less effective than surgical mask, they were still markedly superior to no mask.  A clinical trial in healthcare workers \cite{MacIntyre2015} showed relatively poor performance for cloth masks relative to medical masks.

Mathematical modeling has been influential in providing deeper understanding on the transmission mechanisms and burden of the ongoing COVID-19 pandemic, contributing to the development of public health policy and understanding.  Most mathematical models of the COVID-19 pandemic can broadly be divided into either population-based, SIR (Kermack-McKendrick)-type models, driven by (potentially stochastic) differential equations \cite{Li2020,Wu2020,Tang2020,Kucharski2020,Calafiore2020,Simha2020,Dehning2020,Nesteruk2020,Zhang2020,Anastassopoulou2020,Moore2020}, or agent-based models \cite{Ferguson2020,Wilder2020,Biswas2020,Chang2020,Estrada2020}, in which individuals typically interact on a network structure and exchange infection stochastically. One difficulty of the latter approach is that the network structure is time-varying and can be difficult, if not impossible, to construct with accuracy. Population-based models, alternatively, may risk being too coarse to capture certain real-world complexities. Many of these models, of course, incorporate features from both paradigms, and the right combination of dynamical, stochastic, data-driven, and network-based methods will always depend on the question of interest.

In \cite{Li2020}, Li et al. imposed a metapopulation structure onto an SEIR-model to account for travel between major cities in China. Notably, they include compartments for both documented and undocumented infections.  Their model suggests that as many as 86\% of all cases went undetected in Wuhan before travel restrictions took effect on January 23, 2020. They additionally estimated that, on a per person basis, asymptomatic individuals were only 55\% as contagious, yet were responsible for 79\% of new infections, given their increased prevalence.  The importance of accounting for asymptomatic individuals has been confirmed by other studies (\cite{Ferguson2020}, \cite{Calafiore2020}). In their model-based assessment of case-fatality ratios, Verity et al. \cite{Verity2020} estimated  that 40--50\% of cases went unidentified in China, as of February 8, 2020, while in the case of the Princess Diamond cruise ship, 46.5\% of individuals who tested positive for COVID-19 were asymptomatic \cite{Moriarty2020}.  Further, Calafiore et al.~\cite{Calafiore2020}, using a modified SIR-model, estimated that, on average, cases in Italy went underreported by a factor of 63, as of March 30, 2020.

Several prior mathematical models, motivated by the potential for pandemic influenza, have examined the utility of mask wearing by the general public.  These include a relatively simple modification of an SIR-type model by Brienen et al. \cite{Brienen2010}, while Tracht et al. \cite{Tracht2010} considered a more complex SEIR model that explicitly disaggregated those that do and do not use masks.  The latter concluded that, for pandemic H1N1 influenza, modestly effective masks (20\%) could halve total infections, while if masks were just 50\% effective as source control, the epidemic could be essentially eliminated if just 25\% of the population wore masks.

We adapt these previously developed SEIR model frameworks for transmission dynamics to explore the potential community-wide impact of public use of face masks, of varying efficacy and compliance, on the transmission dynamics and control of the COVID-19 pandemic.  In particular, we develop a two-group model, which stratifies the total population into those who habitually do and do not wear face masks in public or other settings where transmission may occur.  This model takes the form of a deterministic system of nonlinear differential equations, and explicitly includes asymptomatically-infectious humans.  We examine mask effectiveness and coverage (i.e., fraction of the population that habitually wears masks) as our two primary parameters of interest.

We explore possible nonlinearities in mask coverage and effectiveness and the interaction of these two parameters; we find that the product of mask effectiveness and coverage level strongly predicts the effect of mask use on epidemiologic outcomes.  Thus, homemade cloth masks are best deployed \textit{en masse} to benefit the population at large.  There is also a potentially strong nonlinear effect of mask use on epidemiologic outcomes of cumulative death and peak hospitalizations.  We note a possible temporal effect: Delaying mass mask adoption too long may undermine its efficacy.  Moreover, we perform simulated case studies using mortality data for New York and Washington state.  These case studies likewise suggest a beneficial role to mass adoption of even poorly effective masks, with the relative benefit likely greater in Washington state, where baseline transmission is less intense.  The absolute potential for saving lives is still, however, greater under the more intense transmission dynamics in New York state.  Thus, early adoption of masks is useful regardless of transmission intensities, and should not be delayed even if the case load/mortality seems relatively low.

In summary, the benefit to routine face mask use by the general public during the COVID-19 pandemic remains uncertain, but our initial mathematical modeling work suggests a possible strong potential benefit to near universal adoption of even weakly effective homemade masks that may \textit{synergize with}, not replace, other control and mitigation measures.

%

\section{Methods}

\subsection{Baseline mathematical models}

\subsubsection{Model with no mask use}

We consider a baseline model without any mask use to form the foundation for parameter estimation and to estimate transmission rates in New York and Washington state; we also use this model to determine the equivalent transmission rate reductions resulting from public mask use in the full model.

We use a deterministic susceptible, exposed, symptomatic infectious, hospitalized, asymptomatic infectious, and recovered modeling framework, with these classes respectively denoted $S(t)$, $E(t)$, $I(t)$, $H(t)$, $A(t)$, and $R(t)$; we also include $D(t)$ to track cumulative deaths.  We assume that some fraction of detected infectious individuals progress to the hospitalized class, $H(t)$, where they are unable to pass the disease to the general public; we suppose that some fraction of hospitalized patients ultimately require critical care (and may die) \cite{Zhou2020}, but do not explicitly disaggregate, for example, ICU and non-ICU patients.  Based on these assumptions and simplifications, the basic model for the transmission dynamics of COVID-19 is given by the following deterministic system of nonlinear differential equations:

\begin{eqnarray}
\frac{dS}{dt} &=& -\beta(t) (I + \eta A) \frac{S}{N}, \\
\frac{dE}{dt} &=& \beta(t) (I + \eta A) \frac{S}{N} - \sigma E, \\
\frac{dI}{dt} &=& \alpha \sigma E - \phi I - \gamma_I I, \\
\frac{dA}{dt} &=& (1 - \alpha) \sigma E - \gamma_A A, \\
\frac{dH}{dt} &=& \phi I - \delta H - \gamma_H H, \\
\frac{dR}{dt} &=& \gamma_I I + \gamma_A A + \gamma_H H, \\
\frac{dD}{dt} &=& \delta H,
\end{eqnarray}
where
\begin{equation}
N = S + E + I + A + R,
\end{equation}
is the total population in the community, and $\beta(t)$ is the baseline infectious contact rate, which is assumed to vary with time in general, but typically taken fixed.  Additionally, $\eta$ accounts for the relative infectiousness of asymptomatic carriers (in comparison to symptomatic carriers), $\sigma$ is the transition rate from the exposed to infectious class (so $1/\sigma$ is the disease incubation period), $\alpha$ is the fraction of cases that are symptomatic, $\phi$ is the rate at which symptomatic individuals are hospitalized, $\delta$ is the disease-induced death rate, and $\gamma_A$, $\gamma_I$ and $\gamma_H$ are recovery rates for the subscripted population.

We suppose hospitalized persons are not exposed to the general population.  Thus, they are excluded from the tabulation of $N$, and do not contribute to infection rates in the general community.  This general modeling framework is similar to a variety of SEIR-style models recently employed in \cite{Li2020,Ferguson2020}, for example.

For most results in this paper, we use let $\beta(t) \equiv \beta_0$.  However, given ongoing responses to the COVID-19 pandemic in terms of voluntary and mandated social distancing, etc., we also consider the possibility that $\beta$ varies with time and adopt the following functional form from Tang et al. \cite{Tang2020}, with the modification that contact rates do not begin declining from the initial contact rate, $\beta_0$, until time $t_0$,
\begin{equation}
\beta(t) = \left\{
    \begin{array}{lr}
        \beta_0, & t < t_0 \\
        \beta_{min} + (\beta_0 - \beta_{min}) \exp(-r (t - t_0)), & t \geq t_0
    \end{array}
    \right.
\end{equation}
where $\beta_{min}$ is the minimum contact rate and $r$ is the rate at which contact decreases.

\subsection{Baseline epidemiological parameters}

\begin{table}
\footnotesize
\centering
\setlength{\extrarowheight}{4pt}
\begin{tabular}{| l | c | c | c |}  
\hline

\rowcolor[rgb]{.9 .7 .7}
\textbf{Parameter} & \textbf{Likely range (references)} & \textbf{Default value} \\
\hline

$\beta$ (infectious contact rate) & 0.5--1.5 day$^{-1}$ \cite{Shen2020,Read2020,Li2020}, this work & 0.5 day$^{-1}$ \\

\rowcolor[rgb]{.9, .9, .9} $\sigma$ (transition exposed to infectious) & 1/14--1/3 day$^{-1}$ \cite{Lauer2020,Li2020} & 1/5.1 day$^{-1}$ \\

$\eta$ (infectiousness factor for asymptomatic carriers) & 0.4--0.6 \cite{Ferguson2020,Li2020} & 0.5 \\

\rowcolor[rgb]{.9, .9, .9} $\alpha$ (fraction of infections that become symptomatic) & 0.15--0.7 \cite{Li2020,Ferguson2020,Verity2020,Moriarty2020}& 0.5 \\

$\phi$ (rate of hospitalization) & 0.02--0.1 \cite{Zhou2020,Ferguson2020}& 0.025 day$^{-1}$ \\

\rowcolor[rgb]{.9, .9, .9} $\gamma_A$ (recovery rate, Asymptomatic) & 1/14-1/3 day$^{-1}$ \cite{Tang2020,Zhou2020} & 1/7 day$^{-1}$ \\

$\gamma_I$ (recovery rate, symptomatic) & 1/30-1/3 day$^{-1}$ \cite{Tang2020,Zhou2020} & 1/7 day$^{-1}$ \\

\rowcolor[rgb]{.9, .9, .9} $\gamma_H$ (recovery rate, hospitalized) \cite{Tang2020,Zhou2020} & 1/30-1/3 day$^{-1}$ & 1/14 day$^{-1}$ \\

$\delta$ (death rate, hospitalized) & 0.001--0.1 \cite{Ferguson2020} & 0.015 day$^{-1}$ \\

\hline
\end{tabular}

\caption{Baseline model parameters with brief description, likely ranges based on modeling and clinical studies (see text for further details), and default value chosen for this study.}
\label{table:parameters}
\end{table}

\normalsize

The incubation period for COVID-19 is estimated to average 5.1 days \cite{Lauer2020}, similar to other model-based estimates \cite{Li2020}, giving $\sigma = 1/5.1$ day$^{-1}$.  Some previous model-based estimates of infectious duration are on the order of several days \cite{Li2020,Ferguson2020,Tang2020}, with \cite{Tang2020} giving about 7 days for asymptomatic individuals to recover.  However, the clinical course of the disease is typically much longer: In a study of hospitalized patients \cite{Zhou2020}, average total duration of illness until hospital discharge or death was 21 days, and moreover, the median duration of viral shedding was 20 days in survivors.  

The effective transmission rate (as a constant), $\beta_0$, ranges from around 0.5 to 1.5 day$^{-1}$ in prior modeling studies \cite{Read2020,Shen2020,Li2020}, and typically trends down with time \cite{Tang2020,Li2020}.  We have left this as a free parameter in our fits to Washington and New York state mortality data, and find $\beta_0 \approx 0.5$ and $\beta_0 \approx 1.4$ day$^{-1}$ for these states, respectively, values this range.

The relative infectiousness of asymptomatic carriers, $\eta$, is not known, although Ferguson et al.~\cite{Ferguson2020} estimated this parameter at about 0.5, and Li et al.~\cite{Li2020} gave values of 0.42--0.55.  The fraction of cases that are symptomatic, $\alpha$, is also uncertain, with Li et al.~\cite{Li2020} suggesting an overall case reporting rate of just 14\% early in the outbreak in China, but increasing to 65--69\% later; further, $\alpha = 2/3$ was used in \cite{Ferguson2020}.  In the case of the Diamond Princess Cruise ship \cite{Moriarty2020}, 712 (19.2\%) passengers and crews tested positive for SARS-CoV-2, with 331 (46.5\%) asymptomatic at the time of testing.  Therefore, we choose $\alpha = 0.5$ as our default.

Given an average time from symptom onset to dyspnea of 7 days in \cite{Zhou2020}, and 9 days to sepsis, a range of 1--10 days to hospitalization, a midpoint of 5 days seems reasonable (see also \cite{Ferguson2020}); $\phi \approx 0.025$ day$^{-1}$ is consistent with  on the order of 5--15\% of symptomatic patients being hospitalized.  If about 15\% of hospitalized patients die \cite{Ferguson2020}, then $\delta \approx 0.015$ day$^{-1}$ (based on $\gamma_H = 1/14$ day$^{-1}$).

\subsubsection{Model with general mask use}

We assume that some fraction of the general population wears masks with uniform inward efficiency (i.e., primary protection against catching disease) of $\epsilon_i$, and outward efficiency (i.e., source control/protection against transmitting disease) of $\epsilon_o$.  We disaggregate all population variables into those that typically do and do not wear masks, respectively subscripted with $U$ and $M$.  Based on the above assumptions and simplifications, the extended multi-group model for COVID-19 (where members of the general public wear masks in public) is given by:

\small
\begin{eqnarray}
\frac{dS_U}{dt} &=& -\beta (I_U + \eta A_U) \frac{S_U}{N} - \beta \bigl( (1-\epsilon_o) I_M + (1-\epsilon_o) \eta A_M \bigr) \frac{S_U}{N}, \\
\frac{dE_U}{dt} &=& \beta (I_U + \eta A_U) \frac{S_U}{N} + \beta ((1-\epsilon_o) I_M + (1-\epsilon_o) \eta A_M) \frac{S_U}{N} - \sigma E_U, \\
\frac{dI_U}{dt} &=& \alpha \sigma E_U - \phi I_U - \gamma_I I_U, \\
\frac{dA_U}{dt} &=& (1 - \alpha) \sigma E_U - \gamma_A A_U, \\
\frac{dH_U}{dt} &=& \phi I_U - \delta H_U - \gamma_H H_U, \\
\frac{dR_U}{dt} &=& \gamma_I I_U + \gamma_A A_U + \gamma_H H_U, \\
\frac{dD_U}{dt} &=& \delta H_U, \\
\frac{dS_M}{dt} &=& -\beta (1 - \epsilon_i) (I_U + \eta A_U) \frac{S_M}{N} - \beta (1 - \epsilon_i) ((1-\epsilon_o) I_M + (1-\epsilon_o) \eta A_M) \frac{S_M}{N}, \\
\frac{dE_M}{dt} &=& \beta (1 - \epsilon_i) (I_U + \eta A_U) \frac{S_M}{N} + \beta (1 - \epsilon_i) ((1-\epsilon_o) I_M + (1-\epsilon_o) \eta A_M) \frac{S_M}{N} - \sigma E_M, \\
\frac{dI_M}{dt} &=& \alpha \sigma E_M - \phi I_M - \gamma_I I_M, \\
\frac{dA_M}{dt} &=& (1 - \alpha) \sigma E_M - \gamma_A A_M, \\
\frac{dH_M}{dt} &=& \phi I_M - \delta H_M - \gamma_H H_M, \\
\frac{dR_M}{dt} &=& \gamma_I I_M + \gamma_A A_M + \gamma_H H_M, \\
\frac{dD_M}{dt} &=& \delta H_M, \\
\end{eqnarray}
\normalsize
where
\begin{equation}
N = S_U + E_U + I_U + A_U + R_U + S_M + E_M + I_M + A_M + R_M.
\end{equation}

While much more complex than the baseline model, most of the complexity lies in what are essentially bookkeeping terms.  We also consider a reduced version of the above model (equations not shown), such that only symptomatically infected persons wear a mask, to compare the consequences of the common recommendation that only those experiencing symptoms (and their immediate caretakers) wear masks with more general population coverage.

\subsection{Mask efficiency parameters}

We assume a roughly linear relationship between the overall filtering efficiency of a mask and clinical efficiency in terms of either inward efficiency (i.e., effect on $\epsilon_i$) or outward efficiency ($\epsilon_o$), based on \cite{Brienen2010}.  The fit factor for homemade masks averaged 2 in \cite{Davies2013}, while the fit factor averaged 5 for surgical masks.  When volunteers coughed into a mask, depending upon sampling method, the number of colony-forming units resulting varied from 17\% to 50\% for homemade masks and 0--30\% for surgical masks, relative to no mask \cite{Davies2013}.

Surgical masks reduced \textit{P. aeruginosa} infected aerosols produced by coughing by over 80\% in cystic fibrosis patients in \cite{Driessche2015}, while surgical masks reduced CFU count by $>$90\% in a similar study \cite{Stockwell2018}.  N95 masks were more effective in both studies.  Homemade teacloth masks had an inward efficiency between 58 and 77\% over 3 hours of wear in \cite{Sande2008}, while inward efficiency ranged 72--85\% and 98--99\% for surgical and N95-equivalent masks.  Outward efficiency was marginal for teacloth masks, and about 50--70\% for medical masks.  Surgical masks worn by tuberculosis patients also reduced the infectiousness of hospital ward air in \cite{Dharmadhikari2012}, and Leung et al. \cite{Leung2020} very recently observed surgical masks to decrease infectious aerosol produced by individuals with seasonal coronaviruses.  Manikin studies seem to recommend masks as especially valuable under coughing conditions for both source control \cite{Patel2016} and prevention \cite{Lai2012}.  

We therefore estimate that inward mask efficiency could range widely, anywhere from 20--80\% for cloth masks, with $\geq$50\% possibly more typical (and higher values are possible for well-made, tightly fitting masks made of optimal materials), 70--90\% typical for surgical masks, and $>$95\% typical for properly worn N95 masks.  Outward mask efficiency could range from practically zero to over 80\% for homemade masks, with 50\% perhaps typical, while surgical masks and N95 masks are likely 50--90\% and 70--100\% outwardly protective, respectively.

\subsection{Data and model fitting}

We use state-level time series for cumulative mortality data compiled by Center for Systems Science and Engineering at Johns Hopkins University \cite{data}, from January 22, 2020, through April 2, 2020, to calibrate the model initial conditions and infective contact rate, $\beta_0$, as well as $\beta_{min}$ when $\beta(t)$ is taken as an explicit function of time.  Other parameters are fixed at default values in Table \ref{table:parameters}.  Parameter fitting was performed using nonlinear least squares algorithm implemented using the \verb|lsqnonlin| function in MATLAB.  We consider two US states in particular as case studies, New York and Washington, and total population data for each state was defined according to US Census data for July 1, 2019 \cite{Census}.



\section{Results}
%
%

\subsection{Analytic results}

Closed-form expressions for the basic reproduction number, $\mathcal{R}_0$, for the baseline model without masks and the full model with masks are given, for $\beta(t) \equiv \beta_0$, in Appendix A and B, respectively.

\subsection{Masks coverage/efficacy/time to adoption in simulated epidemics}

\subsubsection{Mask/efficacy interaction under immediate adoption}

\begin{figure}
\centering
\includegraphics[scale=.4]{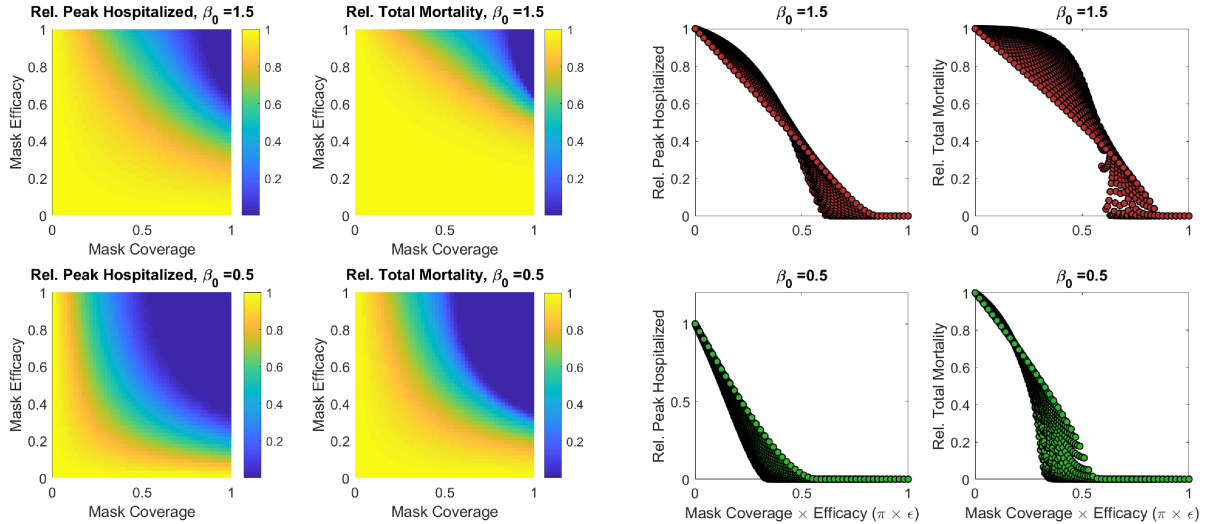}

\caption{Relative peak hospitalizations and cumulative mortality under simulated epidemics, under either a base $\beta_0$ = 0.5 or 1.5 day $^{-1}$, under different general mask coverage level and efficacies (where $\epsilon_o = \epsilon_i = \epsilon$).  Results are relative to a base case with no mask use.  The left half of the figure gives these metrics as two-dimensional functions of coverage and efficacy.  The right half gives these metrics as one-dimensional functions of coverage $\times$ efficacy.}

\label{fig:base_results}
\end{figure}

We run simulated epidemics using either $\beta_0$ = 0.5 or 1.5 day $^{-1}$, with other parameters set to the defaults given in Table \ref{table:parameters}.  These parameter sets give epidemic doubling times early in time (in terms of cumulative cases and deaths) of approximately seven or three days, respectively, corresponding to case and mortality doubling times observed (early in time) in Washington and New York state, respectively.  We use as initial conditions a normalized population of 1 million persons, all of whom are initially susceptible, except 50 initially symptomatically infected (i.e., 5 out 100,000 is the initial infection rate), not wearing masks.

We choose some fraction of the population to be initially in the masked class (``mask coverage''), which we also denote $\pi$, and assume $\epsilon_o = \epsilon_i = \epsilon$.  The epidemic is allowed to run its course (18 simulated months) under constant conditions, and the outcomes of interest are peak hospitalization, cumulative deaths, and total recovered.  These results are normalized against the counterfactual of no mask coverage, and results are presented as heat maps in Figure \ref{fig:base_results}.

Note that the product $\epsilon \times \pi$ predicts quite well the effect of mask deployment: Figure \ref{fig:base_results} also shows (relative) peak hospitalizations and cumulative deaths as functions of this product.  There is, however, a slight asymmetry between coverage and efficacy, such that increasing coverage of moderately effective masks is generally more useful than increasing the effectiveness of masks from a starting point of moderate coverage.

\subsubsection{Delayed adoption}

We run the simulated epidemics described, supposing the entire population is unmasked until mass mask adoption after some discrete delay.  The level of adoption is also fixed as a constant.  We find that a small delay in mask adoption (without any changes in $\beta$) has little effect on peak hospitalized fraction or cumulative deaths, but the ``point of no return'' can rapidly be crossed, if mask adoption is delayed until near the time at which the epidemic otherwise crests.  This general pattern holds regardless of $\beta_0$, but the point of no return is further in the future for smaller $\beta_0$.

\subsection{Mask use and equivalent $\beta$ reduction}

\begin{figure}
\centering
\includegraphics[scale=.425]{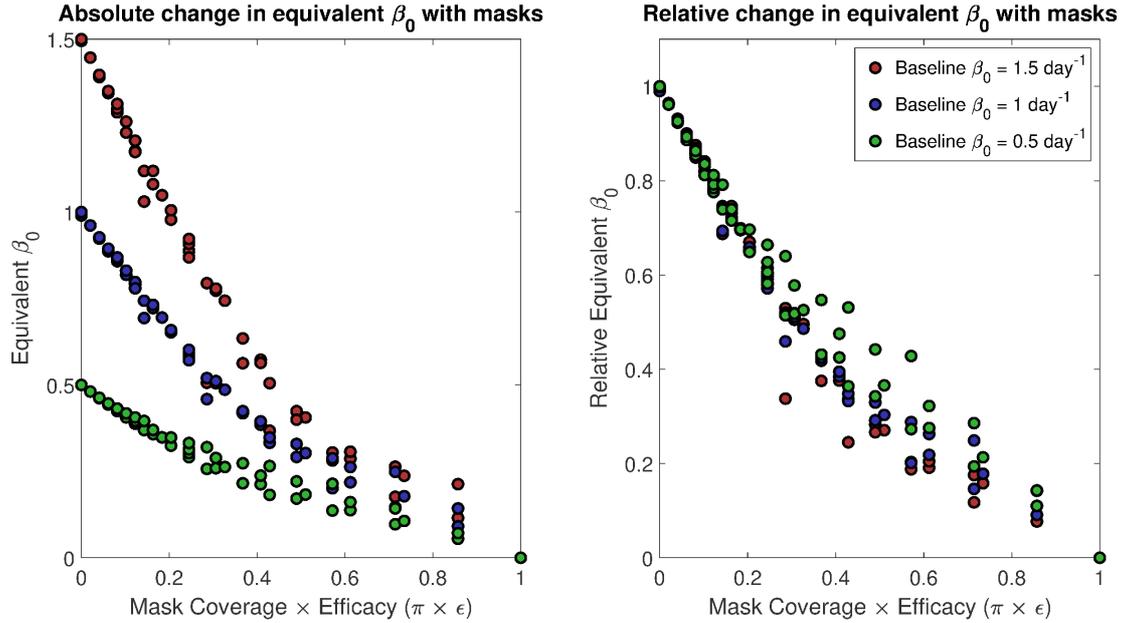}

\caption{Equivalent $\beta_0$, $\tilde{\beta}_0$ (infectious contact rate) under baseline model dynamics as a function of mask coverage $\times$ efficacy, with the left panel giving the absolute value, and the right giving the ratio of $\tilde{\beta}_0$ to the true $\beta_0$ in the simulation with masks.  That is, simulated epidemics are run with mask coverage and effectiveness ranging from 0 to 1, and the outcomes are tracked as synthetic data.  The baseline model without mask dynamics is then fit to this synthetic data, with $\beta_0$ the trainable parameter; the resulting $\beta_0$ is the $\tilde{\beta}_0$.  This is done for simulated epidemics with a true $\beta_0$ of 1.5, 1, or 0.5 day$^{-1}$.}

\label{fig:equivalent_beta}
\end{figure}

The relationship between mask coverage, efficacy, and metrics of epidemic severity considered above are highly nonlinear.  The relationship between $\beta_0$ (the infectious contact rate) and such metrics is similarly nonlinear.  However, incremental reductions in $\beta_0$, due to social distances measures, etc., can ultimately synergize with other reductions to yield a meaningfully effect on the epidemic.  Therefore, we numerically determine what the \textit{equivalent} change in $\beta_0$ under the baseline would have been under mask use at different coverage/efficacy levels, and we denote the equivalent $\beta_0$ value as $\tilde{\beta}_0$.

That is, we numerically simulate an epidemic with and without masks, with a fixed $\beta_0$.  Then, we fit the baseline model to this (simulated) case data, yielding a new \textit{equivalent} $\beta_0$, $\tilde{\beta}_0$.  An excellent fit giving $\tilde{\beta}_0$ can almost always be obtained, though occasionally results are extremely sensitive to $\beta_0$ for high mask coverage/efficacy, yielding somewhat poorer fits.  Results are summarized in Figure \ref{fig:equivalent_beta}, where the $\tilde{\beta}_0$ values obtained and the relative changes in equivalent $\beta$ (i.e., ($\tilde{\beta}_0$) / ($\beta_0$)) are plotted as functions of efficacy times coverage, $\epsilon \times \pi$, under simulated epidemics with three baseline (true) $\beta_0$ values.

From Figure \ref{fig:equivalent_beta}, we see that even 50\% coverage with 50\% effective masks roughly halves the effective disease transmission rate.  Widespread adoption, say 80\% coverage, of masks that are only 20\% effective still reduces the effective transmission rate by about one-third.


\subsection{Outward vs. inward efficiency}

\begin{figure}
\centering
\includegraphics[scale=.425]{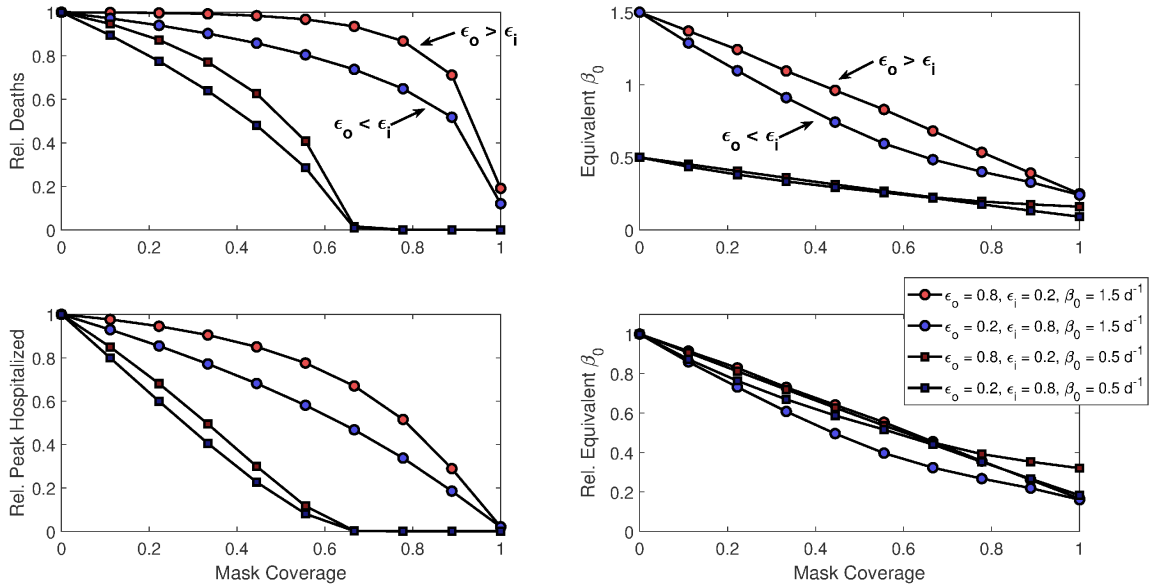}

\caption{Epidemiologic outcomes and equivalent $\beta_0$ changes as a function of mask coverage when masks are either much better at blocking outgoing ($\epsilon_o = 0.8$, $\epsilon_i = 0.2$)  or incoming ($\epsilon_0 = 0.2$, $\epsilon_i = 0.8$) transmission.  Results are demonstrated for both mask permutations under simulated epidemics with baseline $\beta_0$ = 0.5 or 1.5 day $^{-1}$.}

\label{fig:mask_asymmetry}
\end{figure}

Figure \ref{fig:mask_asymmetry} demonstrates the effect of mask coverage on peak hospitalizations, cumulative deaths, and equivalent $\beta_0$ values when either $\epsilon_o = 0.2$ and $\epsilon_i = 0.8$, or visa versa (and for simulated epidemics using either $\beta_0$ = 0.5 or 1.5 day $^{-1}$.  These results suggest that, all else equal, the protection masks afford against acquiring infection ($\epsilon_o$) is actually slightly more important than protection against transmitting infection ($\epsilon_i$), although there is overall little meaningful asymmetry.



\subsection{Masks for symptomatic alone vs. general population}

\begin{figure}
\centering
\includegraphics[scale=.425]{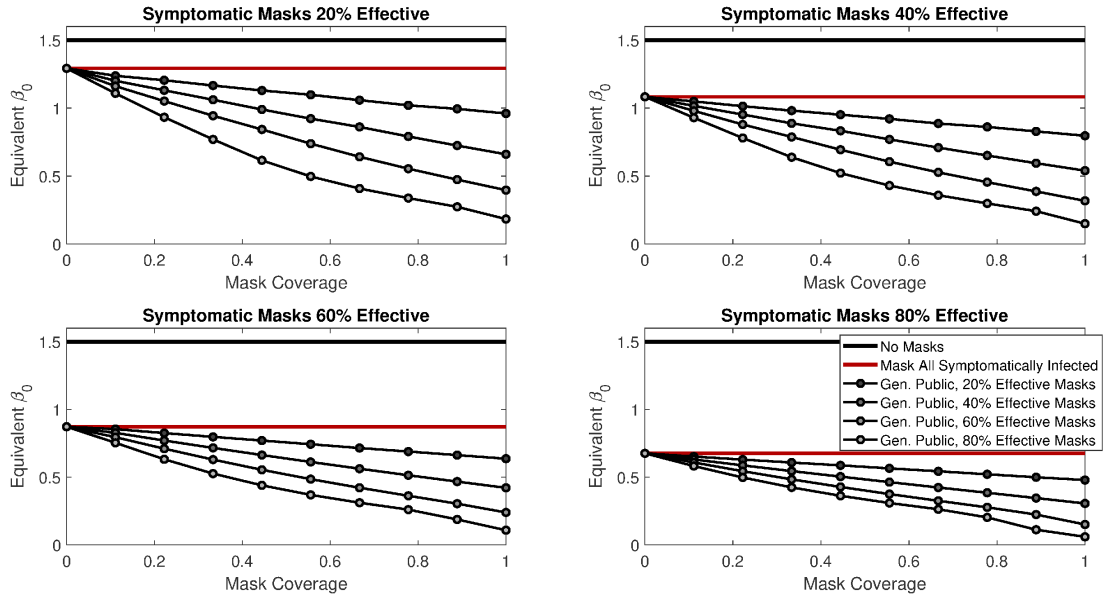}

\caption{Equivalent $\beta_0$ under the model where \textit{all} symptomatic persons wear a mask (whether they otherwise habitually wear a mask or not), under varying levels of effective for the masks given to the symptomatic ($\epsilon_o^I$), and in combination with different degrees of coverage and effectiveness for masks used by the rest of the general public.  Results are for simulated epidemics with a baseline $\beta_0$ of 1.5 day$^{-1}$.}

\label{fig:symptomatic_mask_results}
\end{figure}

Finally, we consider numerical experiments where masks are given to all symptomatically infected persons, whether they otherwise habitually wear masks or not (i.e., both $I_U$ and $I_M$ actually wear masks).  We explore how universal mask use in symptomatically infected persons interacts with mask coverage among the general population; we let $\epsilon_o^I$ represent the effectiveness of masks in the symptomatic, not necessarily equal to $\epsilon_o$.  We again run simulated epidemics with no masks, universal masks among the symptomatic, and then compare different levels of mask coverage in the general (asymptomatic) population.  In this section, we use equivalent $\beta_0$ as our primary metric.  Figure \ref{fig:symptomatic_mask_results} shows how this metric varies as a function of the mask effectiveness given to symptomatic persons, along with the coverage and effectiveness of masks worn by the general public.

We also explore how conclusions vary when either 25\%, 50\%, or 75\% of infectious COVID-19 patients are \textit{asymptomatic} (i.e., we vary $\alpha$).  Unsurprisingly, the greater the proportion of infected people are asymptomatic, the more benefit there is to giving the general public masks in addition to those experiencing symptoms.

\subsection{Simulated case studies: New York \& Washington states}

\begin{figure}
\centering
\includegraphics[scale=.425]{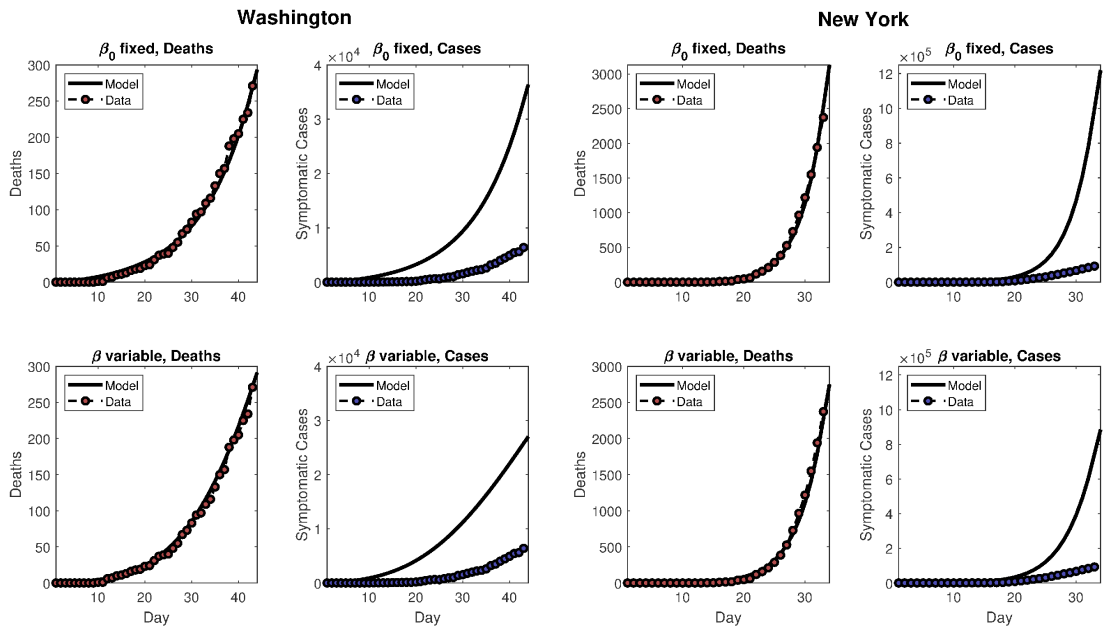}

\caption{The left half of the figure gives model predictions and data for Washington state, using either a constant (top panels) or variable $\beta$ (bottom panel), as described in the test.  The right half of the figure is similar, but for New York state.}
\label{fig:data_fits}
\end{figure}

\begin{figure}
\centering
\includegraphics[scale=.45]{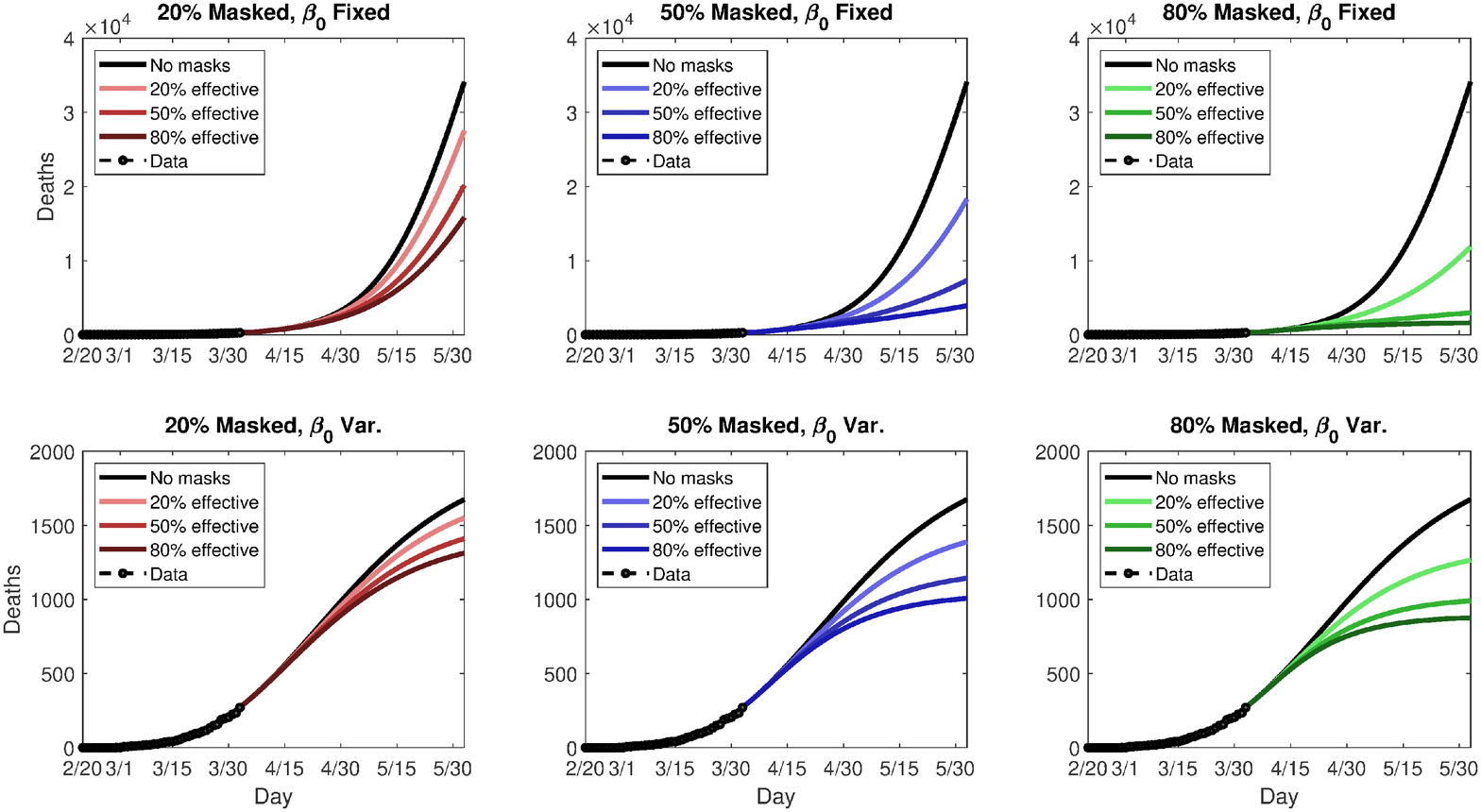}

\caption{Simulated future (cumulative) death tolls for Washington state, using either a fixed (top panels) or variable (bottom panels) transmission rate, $\beta$, and nine different permutations of general public mask coverage and effectiveness.  The y-axes are scaled differently in top and bottom panels.}
\label{fig:sim_washington}
\end{figure}

\begin{figure}
\centering
\includegraphics[scale=.45]{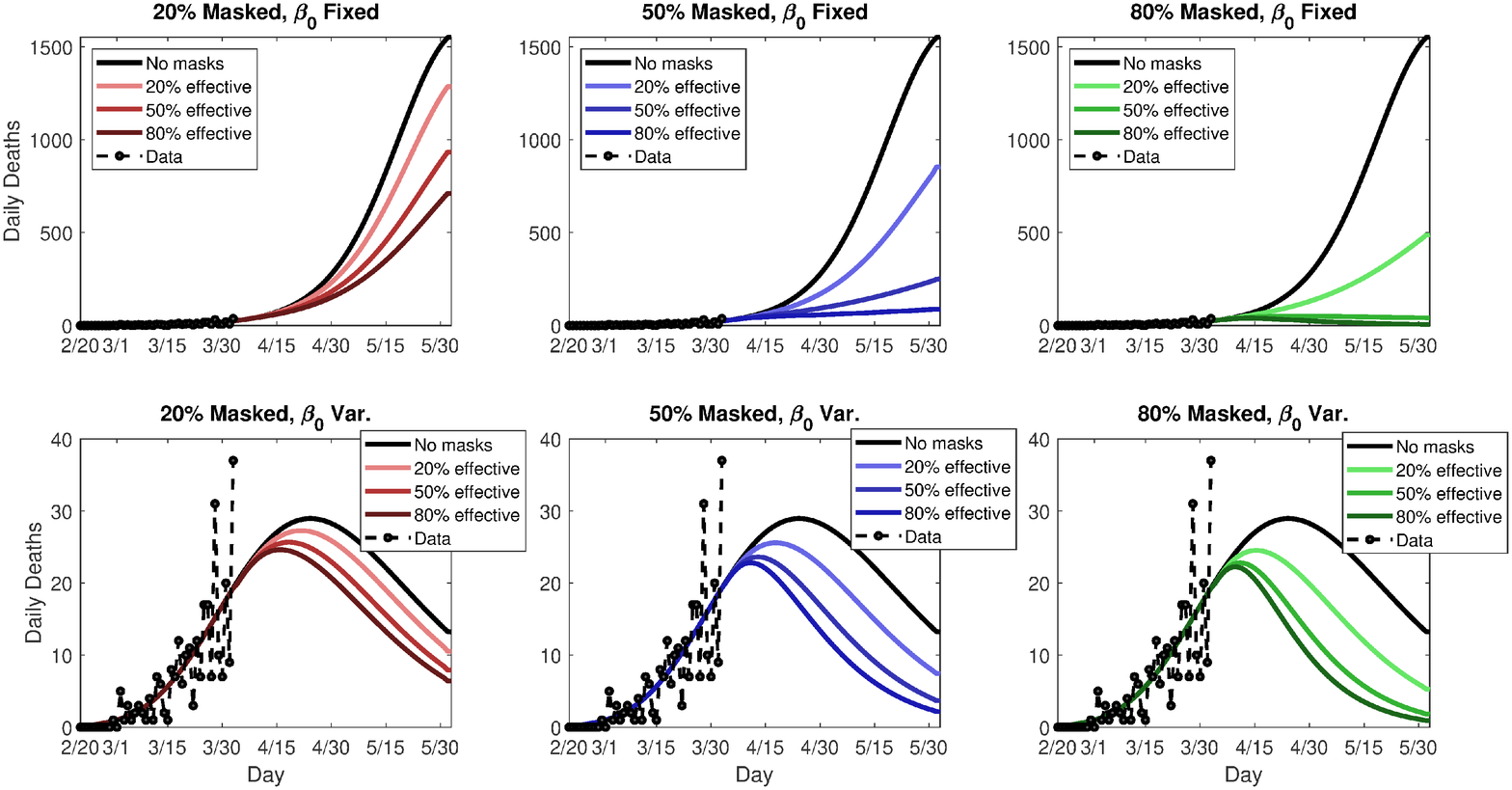}

\caption{Simulated future \textit{daily} death rates for Washington state, using either a fixed (top panels) or variable (bottom panels) transmission rate, $\beta$, and nine different permutations of general public mask coverage and effectiveness.  The y-axes are scaled differently in top and bottom panels.}
\label{fig:sim_washington_daily}
\end{figure}

\begin{figure}
\centering
\includegraphics[scale=.45]{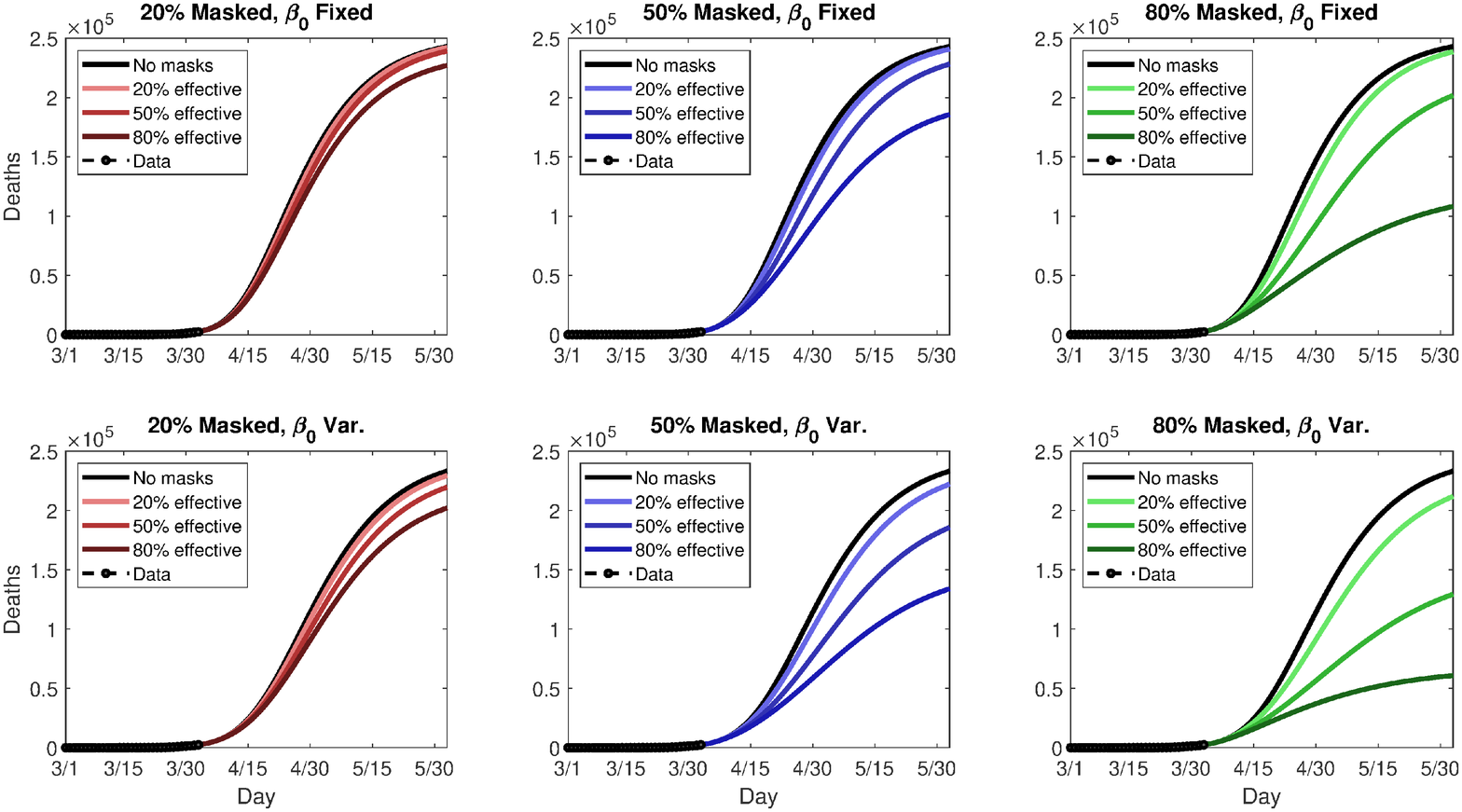}

\caption{Simulated future (cumulative) death tolls for New York state, using either a fixed (top panels) or variable (bottom panels) transmission rate, $\beta$, and nine different permutations of general public mask coverage and effectiveness.}
\label{fig:sim_new_york}
\end{figure}

\begin{figure}
\centering
\includegraphics[scale=.45]{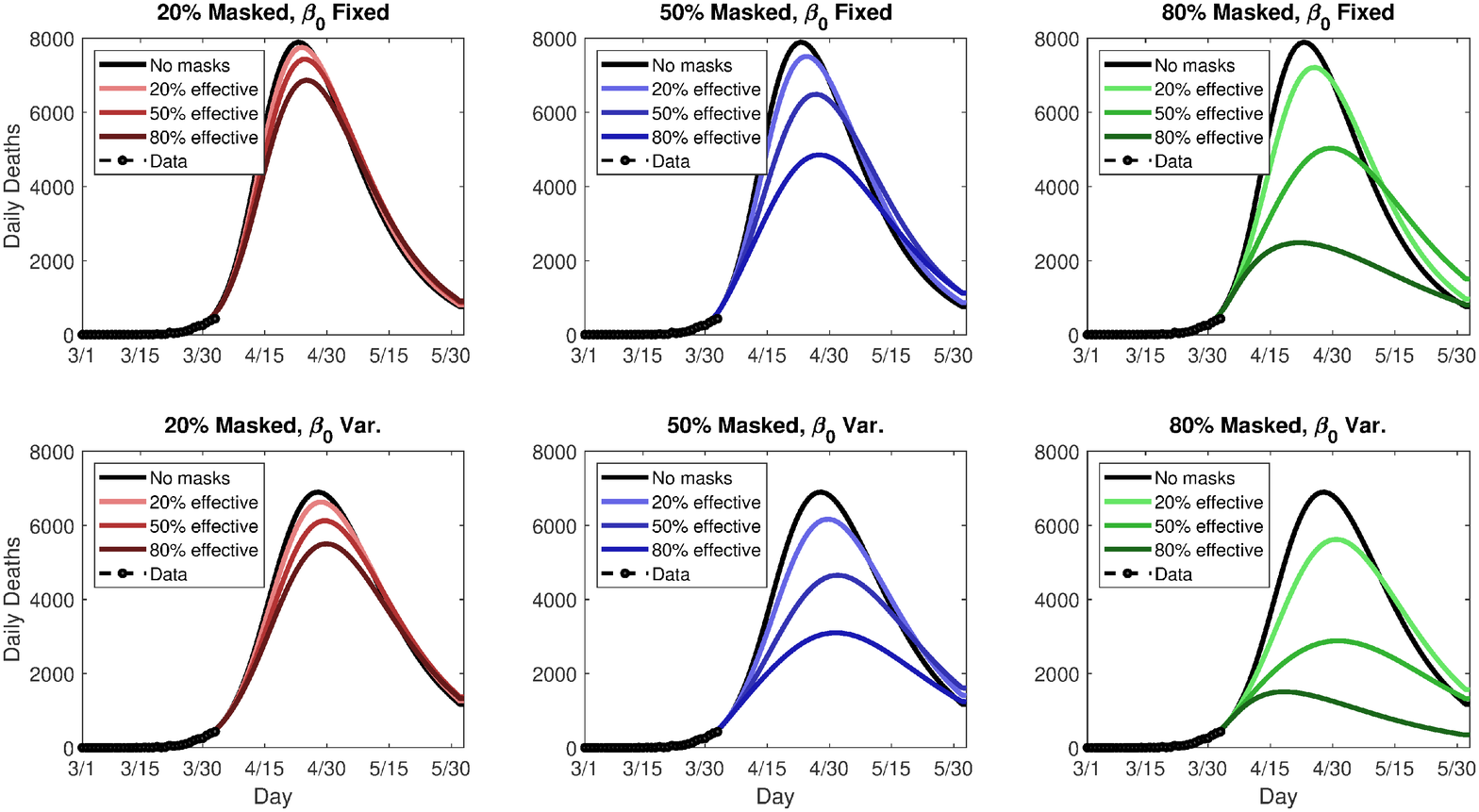}

\caption{Simulated future \textit{daily} death rates for New York state, using either a fixed (top panels) or variable (bottom panels) transmission rate, $\beta$, and nine different permutations of general public mask coverage and effectiveness.}
\label{fig:sim_new_york_daily}
\end{figure}

Fitting to cumulative death data, we use the baseline model to determine the best fixed $\beta_0$ and $I(0)$ for cumulative death data for New York and Washington state.  We use New York state data beginning on March 1, 2020, through April 2, 2020, and Washington state data from February 20, 2020 through April 2, 2020.  For New York state, best-fit parameters are $I(0)$ = 208 (range 154--264) and $\beta_0$ = 1.40 (1.35--1.46) day$^{-1}$ under fixed $\beta_0$.  For the time-varying $\beta(t)$, we fix $r = 0.03$ day$^{-1}$ and $t_0 = 20$, yielding a best-fit $\beta_0$ = 1.33 (1.24--1.42) day$^{-1}$, $\beta_{min}$ = 0.51 (-0.25--1.26) day$^{-1}$, and $I(0)$ = 293 (191--394).

For Washington state, parameters are $I(0)$ = 622 (571--673) and $\beta_0$ = 0.50 (0.49--0.52) day$^{-1}$ under fixed $\beta_0$.  For time-varying $\beta(t)$, we fix $r = 0.04$ day$^{-1}$ and $t_0 = 0$, to yield a best-fit $\beta_0$ = 1.0 (0.87--1.23) day$^{-1}$, $\beta_{min}$ = 0.10 (0--0.19) day$^{-1}$, and $I(0)$ = 238 (177--300).

We fix $r$ and $t_0$, as it is not possible to uniquely identify $r$, $t_0$ and $\beta_{min}$, from death or case data alone (see e.g., \cite{Roda2020} on identifiability problems).  Figure \ref{fig:data_fits} gives cumulative death and case data versus the model predictions for the two states, and for the two choices of $\beta(t)$.  Note that while modeled and actual cumulative deaths match well, model-predicted cases markedly exceed reported cases in the data, consistent with the notion of broad underreporting.

We then consider either fixed $\beta_0$ or time-varying $\beta(t)$, according to the parameters above, in combination with the following \textit{purely hypothetical} scenarios in each state.

\begin{enumerate}
\itemsep0em
\item No masks, epidemic runs its course unaltered with either $\beta(t) \equiv \beta_0$ fixed or $\beta(t)$ variable as described above.
\item The two $\beta$ scenarios are considered in combination with: (1) weak, moderate, or strong deployment of masks, such that $\pi$ = 0.2, 0.5, or 0.8; and (2) weak, moderate, or strong masks, such that $\epsilon$ = 0.2, 0.5, or 0.8.  No masks are used up until April 2, 2020, and then these coverage levels are instantaneously imposed.
\end{enumerate}

This yields 18 scenarios in all (nine mask coverage/efficacy scenarios, plus two underlying trends).  Following the modeled imposition of masks on April 2, 2020, the scenarios are run for 60 additional simulated days.  Figures \ref{fig:sim_washington} and \ref{fig:sim_new_york} summarize the future modeled death toll in each city under the 18 different scenarios, along with historical mortality data.  Figures \ref{fig:sim_washington_daily} and \ref{fig:sim_new_york_daily} show modeled daily death rates, with deaths peaking sometime in late April in New York state under all scenarios, while deaths could peak anywhere from mid-April to later than May, for Washington state.  We emphasize that these are hypothetical and exploratory results, with possible death tolls varying dramatically based upon the future course of $\beta(t)$.  However, the results do suggest that even modestly effective masks, if widely used, could help ``bend the curve,'' with the relative benefit greater in combination with a lower baseline $\beta_0$ or stronger underlying trend towards smaller $\beta(t)$ (i.e., in Washington vs. New York).

\section{Discussion \& Conclusions}

This study aims to contribute to this debate by providing realistic insight into the community-wide impact of widespread use of face masks by members of the general population.
We designed a mathematical model, parameterized using data relevant to COVID-19 transmission dynamics in two US states (New York and Washington).  The model suggests a nontrivial benefit to face mask use by the general public that may vary nonlinearly with mask effectiveness, coverage, and baseline disease transmission intensity.  Face masks should be advised not just for those experiencing symptoms, and likely protect both truly healthy wearers and avoid transmission by asymptomatic carriers.  The community-wide benefits are greatest when mask coverage is as near universal as possible.

There is considerable ongoing debate on whether to recommend general public face mask use (likely mostly homemade cloth masks or other improvised face coverings) \cite{Chan2020}, and while the situation is in flux, more authorities are recommending public mask use, though they continue to (rightly) cite appreciable uncertainty.  With this study, we hope to help inform this debate by providing insight into the potential community-wide impact of widespread face mask use by members of the general population.  We have designed a mathematical model, parameterized using data relevant to COVID-19 transmission dynamics in two US states (New York and Washington), and our model suggests nontrivial and possibly quite strong benefit to general face mask use.  The population-level benefit is greater the earlier masks are adopted, and at least some benefit is realized across a range of epidemic intensities.  Moreover, even if they have, as a sole intervention, little influence on epidemic outcomes, face masks decrease the equivalent effective transmission rate ($\beta_0$ in our model), and thus can stack with other interventions, including social distancing and hygienic measures especially, to ultimately drive nonlinear decreases in epidemic mortality and healthcare system burden.  It bears repeating that our model results are consistent with the idea that face masks, while no panacea, may synergize with other non-pharmaceutical control measures and should be used in combination with and not in lieu of these.

Under simulated epidemics, the effectiveness of face masks in altering the epidemiologic outcomes of peak hospitalization and total deaths is a highly nonlinear function of both mask efficacy and coverage in the population (see Figure \ref{fig:base_results}), with the product of mask efficacy and coverage a good one-dimensional surrogate for the effect.  We have determined how mask use in the full model alters the \textit{equivalent} $\beta_0$, denoted $\tilde{\beta}_0$, under baseline model (without masks), finding this equivalent $\tilde{\beta}_0$ to vary nearly linearly with efficacy $\times$ coverage (Figure \ref{fig:equivalent_beta}).

Masks alone, unless they are highly effective and nearly universal, may have only a small effect (but still nontrivial, in terms of absolute lives saved) in more severe epidemics, such as the ongoing epidemic in New York state.  However, the relative benefit to general masks use may increase with other decreases in $\beta_0$, such that masks can synergize with other public health measures.  Thus, it is important that masks not be viewed as an alternative, but as a complement, to other public health control measures (including non-pharmaceutical interventions, such as social distancing, self-isolation etc.).  Delaying mask adoption is also detrimental.  These factors together indicate that even in areas or states where the COVID-19 burden is low (e.g. the Dakotas), early aggressive action that includes face masks may pay dividends.

These general conclusions are illustrated by our simulated case studies, in which we have tuned the infectious contact rate, $\beta$ (either as fixed $\beta_0$ or time-varying $\beta(t)$), to cumulative mortality data for Washington and New York state through April 2, 2020, and imposed hypothetical mask adoption scenarios.  The estimated range for $\beta$ is much smaller in Washington state, consistent with this state's much slower epidemic growth rate and doubling time.  Model fitting also suggests that total symptomatic cases may be dramatically undercounted in both areas, consistent with prior conclusions on the pandemic \cite{Li2020}.  Simulated futures for both states suggest that broad adoption of even weak masks use could help avoid many deaths, but the greatest relative death reductions are generally seen when the underlying transmission rate also falls or is low at baseline.

Considering a fixed transmission rate, $\beta_0$, 80\% adoption of 20\%, 50\%, and 80\% effective masks reduces cumulative relative (absolute) mortality by 1.8\% (4,419), 17\% (41,317), and 55\% (134,920), respectively, in New York state.  In Washington state, relative (absolute) mortality reductions are dramatic, amounting to 65\% (22,262), 91\% (31,157), and 95\% (32,529).  When $\beta(t)$ varies with time, New York deaths reductions are 9\% (21,315), 45\% (103,860), and 74\% (172,460), while figures for Washington are 24\% (410), 41\% (684), and 48\% (799).  In the latter case, the epidemic peaks soon even without masks.  Thus, a range of outcomes are possible, but both the absolute and relative benefit to weak masks can be quite large; when the relative benefit is small, the absolute benefit in terms of lives is still highly nontrivial.


Most of our model projected mortality numbers for New York and Washington state are quite high (except for variable $\beta(t)$ in Washington), and likely represent worst-case scenarios as they primarily reflect $\beta$ values early in time.  Thus, they may be dramatic overestimates, depending upon these states' populations ongoing responses to the COVID-19 epidemics.  Nevertheless, the estimated transmission values for the two states, under fixed and variable $\beta(t)$ represent a broad range of possible transmission dynamics, are within the range estimated in prior studies \cite{Shen2020,Read2020,Li2020}, and so we may have some confidence in our general conclusions on the possible range of benefits to masks.  Note also that we have restricted our parameter estimation only to initial conditions and transmission parameters, owing to identifiability problems with more complex models and larger parameter groups (see e.g. \cite{Roda2020}).  For example, the same death data may be consistent with either a large $\beta_0$ and low $\delta$ (death rate), or visa versa.

Considering the subproblem of general public mask use in addition to mask use for source control by any (known) symptomatic person, we find that general face mask use is still highly beneficial (see Figure \ref{fig:symptomatic_mask_results}).  Unsurprisingly, this benefit is greater if a larger proportion of infected people are asymptomatic (i.e., $\alpha$ in the model is smaller).  Moreover, it is not the case that masks are helpful exclusively when worn by asymptomatic infectious persons for source control, but provide benefit when worn by (genuinely) healthy people for prevention as well.  Indeed, if there is any asymmetry in outward vs. inward mask effectiveness, inward effectiveness is actually slightly preferred, although the direction of this asymmetry matters little with respect to overall epidemiologic outcomes.  At least one experimental study \cite{Patel2016} does suggest that masks may be superior at source control, especially under coughing conditions vs. normal tidal breathing and so any realized benefit of masks in the population may still be more attributable to source control.

This is somewhat surprising, given that $\epsilon_o$ appears more times than $\epsilon_i$ in the model terms giving the forces of infection, which would suggest outward effectiveness to be of greater import at first glance.  Our conclusion runs counter to the notion that general public masks are primarily useful in preventing asymptomatically wearers from transmitting disease: Masks are valuable as both source control and primary prevention.  This may be important to emphasize, as some people who have self-isolated for prolonged periods may reasonably believe that the chance they are asymptomatically infected is very low and therefore do not need a mask if they venture into public, whereas our results indicate they (and the public at large) still stand to benefit.

Our theoretical results still must be interpreted with caution, owing to a combination of potential high rates of noncompliance with mask use in the community, uncertainty with respect to the intrinsic effectiveness of (especially homemade) masks at blocking respiratory droplets and/or aerosols, and even surprising amounts of uncertainty regarding the basic mechanisms for respiratory infection transmission \cite{MacIntyre2017,Bourouiba2020}.  Several lines of evidence support the notion that masks can interfere with respiratory virus transmission, including clinical trials in healthcare workers \cite{Offeddu2017,MacIntyre2017}, experimental studies as reviewed \cite{Sande2008,Lai2012,Davies2013,Dharmadhikari2012,Patel2016}, and case control data from the 2003 SARS epidemic \cite{Wu2004,Lau2004}.  Given the demonstrated efficacy of medical masks in healthcare workers \cite{Offeddu2017}, and their likely superiority over cloth masks in \cite{MacIntyre2015}, it is clearly essential that healthcare works be prioritized when it comes to the most effective medical mask supply.  Fortunately, our theoretical results suggest significant (but potentially highly variable) value even to low quality masks when used widely in the community.

With social distancing orders in place, essential service providers (such as retail workers, emergency services, law enforcement, etc.) represent a special category of concern, as they represent a largely unavoidable high contact node in transmission networks: Individual public-facing workers may come into contact with hundreds or thousands of people in the course of a day, in relatively close contact (e.g. cashiers).  Such contact likely exposes the workers to many asymptomatic carriers, and they may in turn, if asymptomatic, expose many susceptible members of the general public to potential transmission.  Air exposed to multiple infectious persons (e.g. in grocery stores) could also carry a psuedo-steady load of infectious particles, for which masks would be the only plausible prophylactic\cite{Lai2012}.  Thus, targeted, highly effective mask use by service workers may be reasonable.  We are currently extending the basic model framework presented here to examine this hypothesis.

In conclusion, our findings suggest that face mask use should be as nearly universal (i.e., nation-wide) as possible and implemented without delay, even if most mask are homemade and of relatively low quality.  This measure could contribute greatly to controlling the COVID-19 pandemic, with the benefit greatest in conjunction with other non-pharmaceutical interventions that reduce community transmission.  Despite uncertainty, the potential for benefit, the lack of obvious harm, and the precautionary principle lead us to strongly recommend as close to universal (homemade, unless medical masks can be used without diverting healthcare supply) mask use by the general public as possible.


\section*{Acknowledgements}

One of the authors (ABG) acknowledge the support, in part, of the Simons Foundation (Award $\#$585022) and the National Science Foundation (Award 1917512).

\section*{Appendix A: Basic Reproduction Number for Baseline Model}

The basic reproduction number for both the baseline and the full model is for the special case when $\beta(t) \equiv \beta_0$.
The local stability of the DFE is explored using the next generation operator method \cite{D-H,D-W}. Using the notation in \cite{D-W}, it follows that the matrices $\mathcal{F}$ of new infection terms and $\mathcal{V}$ of the remaining transfer terms
associated with the  version of the model are given, respectively, by

\begin{equation*}
 \mathcal{F}=
\left[
\begin{array}{*{20}c}
 0&\beta_0&\beta_0 \eta\\
 0&0&0\\
 0&0&0\
 \end{array}
\right],
\end{equation*}

\begin{equation*}
\mathcal{V}=
\left[
\begin{array}{*{20}c}
 \sigma&0&0\\
 -\alpha \sigma&(\phi+\gamma_I)&0\\
 -(1-\alpha)\sigma&0&\gamma_A\
 \end{array}
\right].
\end{equation*}

 \vspace{2mm}
 \noindent The basic reproduction number of the  model, denoted by $\mathcal{R}_0$, is given by

\begin{equation}
 \mathcal{ R}_0 = \frac{\beta_0\alpha \sigma}{\sigma(\phi+\gamma_I)}+\frac{\beta_0 \eta(1-\alpha)}{\gamma_A}.
\end{equation}

\section*{Appendix B: Basic Reproduction Number for Full Model}

The local stability of the DFE is explored using the next generation operator method \cite{D-H,D-W}. Using the notation in \cite{D-W}, it follows that the matrices $\mathcal{F}$ of new infection terms and $\mathcal{V}$ of the remaining transfer terms
associated with the  version of the model are given, respectively, by

\begin{equation*}
 \mathcal{F}=
\left[
\begin{array}{*{20}c}
 0&\beta_0&\beta_0 \eta&0&\beta_0(1-\epsilon_o)&\beta_0 (1-\epsilon_o) \eta\\
 0&0&0&0&0&0\\
 0&0&0&0&0&0\\
 0&\beta_0(1-\epsilon_i)&\beta_0 (1-\epsilon_i)\eta&0&\beta_0(1-\epsilon_o)(1-\epsilon_i)&\beta_0(1-\epsilon_o)(1-\epsilon_i) \eta\\
 0&0&0&0&0&0\\
 0&0&0&0&0&0\
 \end{array}
\right],
\end{equation*}

\begin{equation*}
\mathcal{V}=
\left[
\begin{array}{*{20}c}
 \sigma&0&0&0&0&0\\
 -\alpha \sigma&(\phi+\gamma_I)&0&0&0&0\\
 -(1-\alpha)\sigma&0&\gamma_A&0&0&0\\
0&0&0& \sigma&0&0\\
 0&0&0&-\alpha \sigma&(\phi+\gamma_I)&0\\
 0&0&0&-(1-\alpha)\sigma&0&\gamma_A\
 \end{array}
\right].
\end{equation*}

 \vspace{2mm}
 \noindent The basic reproduction number of the  model, denoted by $\mathcal{R}_0$, is given by

\begin{equation}
 \mathcal{ R}_0 = \beta_0 [1+(1-\epsilon_o)(1-\epsilon_i)]\left(\frac{\alpha \sigma}{\sigma(\phi+\gamma_I)}+\frac{ \eta(1-\alpha)}{\gamma_A}\right).
\end{equation}

\end{document}